\documentclass[fleqn]{llncs}
\usepackage{version}
\usepackage{st}

\title{Searching Publications on Software Testing}
\author{C.A. Middelburg}
\institute{Informatics Institute, Faculty of Science,
           University of Amsterdam, \\
           Science Park~904, 1098~XH Amsterdam, the Netherlands \\
           \email{C.A.Middelburg@uva.nl}}

\begin{document}
\maketitle

\begin{abstract}
This note concerns a search for publications in which the pragmatic
concept of a test as conducted in the practice of software testing is
formalized, a theory about software testing based on such a
formalization is presented or it is demonstrated on the basis of such a
theory that there are solid grounds to test software in cases where in
principle other forms of analysis could be used.
This note reports on the way in which the search has been carried out
and the main outcomes of the search.
The message of the note is that the fundamentals of software testing are
not yet complete in some respects.
\end{abstract}


\section{Introduction}
\label{sect-introduction}

In practice, it is virtually unavoidable that software has faults.
This implies that there are normally risks involved in using software.
In order to reduce the risks as much as possible, developed software is
put to tests before it is released.
Software testing definitely tends to form an important part of software
development.
A test as conducted in the practice of software testing is an activity
in which certain code is executed under certain conditions, certain
effects of the execution are observed, and an evaluation of the observed
effects is made.
A sound understanding of this pragmatic concept is a prerequisite for
the advancement of software testing.

To check whether more or less appropriate formalized versions of the
pragmatic concept of a test, as well as theories based on them, already
exist, a search for witnessing publications has been carried out.
An issue of utmost importance is the issue of whether or not there are
solid grounds to test software in cases where in principle other forms
of analysis could be used.
For that reason, the above-mentioned search has been combined with a
search for publications treating this issue.
This note reports on the way in which the search has been carried out
and the main outcomes of the search.
In the concluding remarks, the main outcomes of the search are roughly
summarized in the form of two remarkable observations, which clearly
reflect the fact that the fundamentals of software testing are not yet
complete in some respects.
Therefore, it may well be that these observations will serve as triggers
of future work on the fundamentals of software testing.

\section{The Approach to Carry out the Search}
\label{sect-search}

For the search, Google Scholar and the search engines of relevant
journal collections and bibliographies are available.
The search engines of relevant journal collections and bibliographies
have the option to search for publications in which given terms occur in
their title, abstract or keywords, but none cover all relevant
literature.
Google Scholar does not have this option, i.e.\ only full text search is
possible, but it covers all literature covered by the search engines of
relevant journal collections.

In a preparatory phase of the search, the following specifics were found:
\begin{itemize}
\item
the term ``software testing'' rarely occurs in publications whose
subject is not software testing;
\item
in the early days of software testing, the term ``program testing'' is
often used as a synonym for ``software testing''.
\end{itemize}

This led to the following more or less systematic way to carry out the
search:
\begin{itemize}
\item
first, divide the past from 1951 in periods of five years and search for
relevant publications in each period seperately by means of Google
Scholar;
\item
next, check for each relevant publication found in the previous step
whether it references additional relevant publications;
\item
then, repeat the previous step until no additional relevant publications
are found;
\item
finally, search for relevant publications by means of the search engines
of relevant journal collections and bibliographies.
\end{itemize}

The number of publications found by a full text search with Google
Scholar using the search term
\texttt{"program testing" OR "software testing"} is about 15,000.
Google Scholar shows only the 1,000 best matching publications, but by
searching in periods of five years I could go through the 6,300 best
matching publications.
The last step is meant as a double-check: it should yield very few new
relevant publications.
The following journal collections and bibliographies are considered
relevant: ACM Digital Library, SpringerLink, ScienceDirect, IEEE Xplore,
ACM Guide, DBLP Computer Science Bibliography, and the Collection of
Computer Science Bibliographies.

The publications that were considered relevant to the purpose of the
search in the first step turned out to be strongly connected by
citations.
It proved to be unnecessary to do the second step more than twice.

In a previous search for publications, which is reported on
in~\cite{Mid10a}, the term for the concept concerned was a term that
occurs in extremely many publications virtually unrelated to the
concept.
Consequently, that search had to be carried out in a more sophisticated
way.

\section{The Main Outcomes of the Search}
\label{sect-outcome}

Recall that the publications searched for are publications in which the
pragmatic concept of a test as conducted in the practice of software
testing is formalized, a theory about software testing based on such a
formalization is presented or it is demonstrated on the basis of such a
theory that there are solid grounds to test software in cases where in
principle other forms of analysis could be used.

Bear in mind that a test as conducted in the practice of software
testing is an activity in which certain code is executed under certain
conditions, certain effects of the execution are observed, and an
evaluation of the observed effects is made.
It turns out that there is no electronically available publication on
software testing in which this pragmatic concept of a test is
formalized.

Up to around 1970, the term ``program testing'' is used instead of the
term ``software testing''.
The oldest electronically available publication in which the former term
is used seems to be an article by Perry, namely~\cite{Per52a}, which is
published in 1952.
To our knowledge, the oldest electronically available article on
software testing isan article by Sauder, namely~\cite{Sau62a}, which is
published in 1962.
Relatively in-depth theoretical investigations into various important
aspects of software testing begin around 1975.
The aspects that are investigated include test selection (see e.g.%
~\cite{BEL75a,How75a,Cla76a,How80a,RW85a,ROT89a,WJ91a,OLAA03a,Gau10a}),
test evaluation (see e.g.\ \linebreak[3]
~\cite{BDLS80a,Wey83a,Wey88a,Mil92a,UOH93a,FW94a,FW00a,KCM01a}),
testability (see e.g.~\cite{Wey82a,How85a,VM95a,YL98a}),
testing complexity (see e.g.~\cite{Tai80a,Wey90a,Mye92a}), and
reliability/dependability of tested software
(see e.g.\ \linebreak[2] \cite{BL75a,WW88a,Ham94a,Sch09a}).
The theories about software testing that came into being consolidate
the results of those investigations.
No trace of the pragmatic concept of a test outlined above is found in
those theories.
In other words, they do not provide a sound understanding of the concept
of a test as conducted in the practice of software testing.

Among the theories in question, the theories presented
in~\cite{GG75a,How76a,How78a,Gel78a,WO80a,Gou83a,Bou85a,Ber91a}
are usually considered the most interesting ones.
Programs are identified in those theories with objects of various kinds,
including programs in some programming language, flowcharts, total
functions, structures in the sense of mathematical logic, and even
objects of an arbitrary kind related in specific ways with objects of
two other arbitrary kinds --- with which specifications and tests are
identified.
With the exception of programs in some programming language, these
\mbox{(semi-)}\linebreak[2]formalizations of the pragmatic concept of a
program impede the introduction of suitable formalizations of the
pragmatic concept of a test outlined above.

Several existing theories about software testing are introduced by means
of a brief outline of the pragmatic concept of a test that looks much
like the outline given above, but surprisingly even in those theories no
trace of the concept is found.
It looks as if, at the very outset of the development of existing
theories about software testing, it is immediately forgotten that in
software testing a test involves the execution of code and observation
of effects of the execution.
This is amazing because what is forgotten is exactly what sets software
testing apart from other forms of software analysis.
Recommendation Z.500 of the International Telecommunication
Union~\cite{ITU97a} is one of the few publications that goes more deeply
into what is involved in a test, but this standardization document does
not provide a formalization of the concept.

It turns out that there is no electronically available publication on
software testing that seriously treats the issue of whether or not there
are solid grounds to test software in cases where in principle other
forms of analysis could be used.
If the rationale of software testing is given, it is invariably that
software testing is less expensive, less time-consuming and less
error-prone than any other form of analysis that could be used.
This is however a collection of claims for which not a shred of evidence
is produced in the literature on software testing.
This is amazing because it means that the rationale of software testing
has not yet been settled after fifty years of actual software testing.
Clearly, the lack of a theory about software testing that allows of
setting software testing apart from other forms of software analysis has
impeded a rationale with a sound theoretical basis.

\section{Miscellaneous}
\label{sect-miscellaneous}

In Section~\ref{sect-outcome}, the main outcomes of the search are
outlined.
That outline could raise the question what most publications on software
testing are about.
It turns out that they mainly concern the following:
\begin{itemize}
\item
theories and techniques pertaining to particular aspects of software
testing, such as test selection, test evaluation, testability, testing
complexity, reliability of tested software, compositional testing, and
regression testing;
\item
theories and techniques pertaining to software testing with regard to
particular kinds of behaviour, such as input-output behaviour and
interactive behaviour;
\item
issues related to the testing of software developed according to
particular paradigms, such as object-oriented software, component-based
software and service-oriented software;
\item
designs of, analyses of, and experiences with specific software testing
tools and techniques.
\end{itemize}
It is striking that most of these publications give little insight into
the pragmatic concept of a test as conducted in the practice of software
testing.

With regard to the journals in which theoretical articles on software
testing are published, it is noticeable that from about 1975 a
relatively large part of the theoretical articles on software testing is
published in ``IEEE Transactions on Software Engineering''.
It is visible that since 1991 the journal ``Software Testing,
Verification and Reliability'' is becoming increasingly important.

\section{Concluding remarks}
\label{sect-conclusions}

To check whether more or less appropriate formalized versions of the
pragmatic concept of a test, as well as theories based on them, already
exist, a search for witnessing publications have been carried out.
From the main outcomes of the search, it must be concluded that:
\begin{itemize}
\item
there does not exists a formalized version of the pragmatic concept of a
test as conducted in the practice of software testing;
\item
there does not exist a theory about software testing that provides a
sound understanding of the pragmatic concept of a test;
\item
no attention is paid to giving a rationale of software testing that has
a solid theoretical basis.
\end{itemize}
Some remarkable observations are:
\begin{itemize}
\item
in the development of existing theories about software testing, what
sets software testing apart from other forms of software analysis,
namely that a test involves the execution of code and observation of
effects of the execution, is completely forgotten;
\item
there has not been any serious study of the issue of whether or not
there are solid grounds to test software in cases where in principle
other forms of analysis could be used.
\end{itemize}

\bibliographystyle{splncs03}
\bibliography{ST}

\begin{thebibliography}{10}
\providecommand{\url}[1]{\texttt{#1}}
\providecommand{\urlprefix}{URL }

\bibitem{Ber91a}
Bernot, G.: Testing against formal specifications: A theoretical view. In:
  Abramsky, S., Maibaum, T.S.E. (eds.) TAPSOFT '91. Lecture Notes in Computer
  Science, vol. 494, pp. 99--119. Springer-Verlag (1991)

\bibitem{Bou85a}
Boug{\'e}, L.: A contribution to the theory of program testing. Theoretical
  Computer Science  37(2),  151--181 (1985)

\bibitem{BEL75a}
Boyer, R.S., Elspas, B., Levitt, K.N.: {SELECT} --- a formal system for testing
  and debugging programs by symbolic execution. In: Proceedings of the
  International Conference on Reliable Software. pp. 234--245. ACM Press (1975)

\bibitem{BL75a}
Brown, J.R., Lipow, M.: Testing for software reliability. ACM SIGPLAN Notices
  10(6),  518--527 (1975)

\bibitem{BDLS80a}
Budd, T.A., DeMillo, R.A., Lipton, R.J., Sayward, F.G.: Theoretical and
  empirical studies on using program mutation to test the functional
  correctness of programs. In: POPL '80. pp. 220--233. ACM Press (1980)

\bibitem{Cla76a}
Clarke, L.A.: A system to generate test data and symbolically execute programs.
  IEEE Transactions on Software Engineering  2(3),  215--222 (1976)

\bibitem{FW94a}
Fleyshgakker, V.N., Weiss, S.N.: Efficient mutation analysis: A new approach.
  In: ISSTA '94. pp. 185--195. ACM Press (1994)

\bibitem{FW00a}
Frankl, P.G., Weyuker, E.J.: Testing software to detect and reduce risk.
  Journal of Systems and Software  53(3),  275--286 (2000)

\bibitem{Gau10a}
Gaudel, M.C.: Software testing based on formal specification. In: Borba, P.,
  Cavalcanti, A., Sampaio, A., Woodcock, J. (eds.) Testing Techniques in
  Software Engineering. Lecture Notes in Computer Science, vol. 6153, pp.
  215--242. Springer-Verlag (2010)

\bibitem{Gel78a}
Geller, M.: Test data as an aid in proving program correctness. Communications
  of the ACM  21(5),  368--375 (1978)

\bibitem{GG75a}
Goodenough, J.B., Gerhart, S.L.: Toward a theory of test data selection. ACM
  SIGPLAN Notices  10(6),  493--510 (1975)

\bibitem{Gou83a}
Gourlay, J.S.: A mathematical framework for the investigation of testing. IEEE
  Transactions on Software Engineering  9(6),  686--709 (1983)

\bibitem{Ham94a}
Hamlet, D.: Foundations of software testing: Dependability theory. ACM SIGSOFT
  Software Engineering Notes  19(5),  128--139 (1994)

\bibitem{How75a}
Howden, W.E.: Methodology for the generation of program test data. IEEE
  Transactions on Computers  24(5),  554--560 (1975)

\bibitem{How76a}
Howden, W.E.: Reliability of the path analysis testing strategy. IEEE
  Transactions on Software Engineering  2(3),  208--215 (1976)

\bibitem{How78a}
Howden, W.E.: Algebraic program testing. Acta Informatica  10(1),  53--66
  (1978)

\bibitem{How80a}
Howden, W.E.: Functional program testing. IEEE Transactions on Software
  Engineering  6(2),  162--169 (1980)

\bibitem{How85a}
Howden, W.E.: The theory and practice of functional testing. IEEE Software
  2(5),  6--17 (1985)

\bibitem{ITU97a}
ITU-T: Framework on formal methods in conformance testing. ITU-T Recommendation
  Z.500 (1997)

\bibitem{KCM01a}
Kim, S.W., Clark, J.A., McDermid, J.A.: Investigating the effectiveness of
  object-oriented testing strategies using the mutation method. Software
  Testing, Verification and Reliability  11(4),  207--225 (2001)

\bibitem{Mid10a}
Middelburg, C.A.: Searching publications on operating systems. {\tt
  arXiv:1003.\linebreak[2]5525v1 [cs.OS]} (March 2010)

\bibitem{Mil92a}
Miller, K.W., et~al.: Estimating the probability of failure when testing
  reveals no failures. IEEE Transactions on Software Engineering  18(1),
  33--43 (1992)

\bibitem{Mye92a}
Myers, G.J.: The complexity of software testing. Software Engineering Journal
  7(1),  13--24 (1992)

\bibitem{OLAA03a}
Offutt, J., Liu, S., Abdurazik, A., Ammann, P.: Generating test data from
  state-based specifications. Software Testing, Verification and Reliability
  13(1),  25--53 (2003)

\bibitem{Per52a}
Perry, C.L.: The logical design of the {Oak Ridge} digital computer. In: 1952
  ACM Annual Meeting. pp. 23--27. ACM Press (1952)

\bibitem{RW85a}
Rapps, S., Weyuker, E.J.: Selecting software test data using data flow
  information. IEEE Transactions on Software Engineering  11(4),  367--375
  (1985)

\bibitem{ROT89a}
Richardson, D.J., O'Malley, O., Tittle, C.: Approaches to specification-based
  testing. In: TAV3. pp. 86--96. ACM Press (1989)

\bibitem{Sau62a}
Sauder, R.L.: A general test data generator for {COBOL}. In: AIEE-IRE '62
  (Spring). pp. 317--323. ACM Press (1962)

\bibitem{Sch09a}
Schneidewind, N.: Integrating testing with reliability. Software Testing,
  Verification and Reliability  19(3),  175--198 (2009)

\bibitem{Tai80a}
Tai, K.C.: Program testing complexity and test criteria. IEEE Transactions on
  Software Engineering  6(6),  531--538 (1980)

\bibitem{UOH93a}
Untch, R.H., Offutt, A.J., Harrold, M.J.: Mutation analysis using mutant
  schemata. In: ISSTA '93. pp. 139--148. ACM Press (1993)

\bibitem{VM95a}
Voas, J.M., Miller, K.W.: Software testability: The new verification. IEEE
  Software  12(3),  17--28 (1995)

\bibitem{WW88a}
Weiss, S.N., Weyuker, E.J.: An extended domain-based model of software
  reliability. IEEE Transactions on Software Engineering  14(19),  1512--1524
  (1988)

\bibitem{Wey82a}
Weyuker, E.J.: On testing non-testable programs. Computer Journal  25(4),
  465--470 (1982)

\bibitem{Wey83a}
Weyuker, E.J.: Assessing test data adequacy through program inference. ACM
  Transactions on Programming Languages and Systems  5(4),  641--655 (1983)

\bibitem{Wey88a}
Weyuker, E.J.: The evaluation of program-based software test data adequacy
  criteria. Communications of the ACM  31(6),  668--675 (1988)

\bibitem{Wey90a}
Weyuker, E.J.: The cost of data flow testing: An empirical study. IEEE
  Transactions on Software Engineering  16(2),  121--128 (1990)

\bibitem{WJ91a}
Weyuker, E.J., Jeng, B.: Analyzing partition testing strategies. IEEE
  Transactions on Software Engineering  17(7),  703--711 (1991)

\bibitem{WO80a}
Weyuker, E.J., Ostrand, T.J.: Theories of program testing and the application
  of revealing subdomains. IEEE Transactions on Software Engineering  6(3),
  236--246 (1980)

\bibitem{YL98a}
Yeh, P.L., Lin, J.C.: Software testability measurements derived from data flow
  analysis. In: CSMR '98. pp. 96--103. IEEE Computer Society Press (1998)

\end{thebibliography}


\end{document}